\documentclass[reprint,aps, prm,superscriptaddress,amsmath,amssymb]{revtex4-2}
\usepackage{graphicx}
\usepackage{dcolumn}
\usepackage{bm}
\usepackage[mathlines]{lineno}
\usepackage{etoolbox}
\usepackage{gensymb}
\usepackage{subfigure}
\usepackage{amsmath}
\usepackage{amssymb}
\usepackage{dsfont}
\usepackage{hyperref}
\usepackage{multirow}
\usepackage{rotating}
\usepackage{epstopdf}
\usepackage{color}
\usepackage{wrapfig}
\usepackage{fancyhdr}
\usepackage{lipsum}
\usepackage{tabularx}
\usepackage[capitalize]{cleveref}
\newcommand{\IACS}{\affiliation{School of Physical Sciences, Indian Association for the Cultivation of Science, Jadavpur, Kolkata 700032, India}}

\usepackage[T1]{fontenc}
\DeclareUnicodeCharacter{2212}{\~}
\DeclareUnicodeCharacter{2212}{_}
\DeclareUnicodeCharacter{03B4}{$\delta$}
\usepackage[labelfont=bf,labelsep=quad,justification=justified]{caption}
\begin{document}	
	\preprint{APS}	
	\title{Enhanced coercivity and emergent spin-cluster-glass state in 2D ferromagnetic material, Fe$_3$GeTe$_2$}

	\author{Satyabrata Bera}
    \IACS
	
	\author{Suman Kalyan Pradhan}
\IACS
	\author{Riju Pal}
	\affiliation{
		SN Bose National Centre for Basic sciences, Sector III, Block JD, Salt Lake, Kolkata 700106, India.
	}
	\author{Buddhadeb Pal}
	\affiliation{
		SN Bose National Centre for Basic sciences, Sector III, Block JD, Salt Lake, Kolkata 700106, India.
	}
	\author{Arnab Bera}
    \IACS

	\author{Sk Kalimuddin}
    \IACS
	\author{Manjil Das}
	 \IACS

	\author{Deep Singha Roy}
	 	 \IACS
    
\author{Hasan Afzal }
  \IACS
  
	\author{Atindra Nath Pal}
	\affiliation{
		SN Bose National Centre for Basic sciences, Sector III, Block JD, Salt Lake, Kolkata 700106, India.
	}
	\author{Mintu Mondal}
	\email{sspmm4@iacs.res.in}
    \IACS
	\date{\today}
\begin{abstract}	
Two-dimensional (2D) van der Waals (vdW) magnetic materials with high coercivity and high $T_\text{C}$ are desired for spintronics and memory storage applications. Fe$_3$GeTe$_2$ (F3GT) is one such 2D vdW ferromagnet with a reasonably high $T_\text{C}$,  but with a very low coercive field, $H_\text{c}$  ($\lesssim$100~Oe).  Some of the common techniques of enhancing $H_\text{c}$ are by introducing pinning centers, defects, stress, doping, etc. They involve the risk of undesirable alteration of other important magnetic properties. Here we propose a very easy, robust, and highly effective method of phase engineering by altering the sample growth conditions to greatly enhance the intrinsic coercivity (7-10 times) of the sample, without compromising its fundamental magnetic properties ($T_\text{C}\simeq$210K). The phase-engineered sample (F3GT-2) comprises of parent F3GT phase with a small percentage of randomly embedded clusters of a coplanar FeTe (FT) phase. The FT phase serves as both mosaic pinning centers between grains of F3GT above its antiferromagnetic transition temperature ($T_\text{C1}\sim$70~K) and also as anti-phase domains below $T_\text{C1}$. As a result, the grain boundary disorder and metastable nature are greatly augmented, leading to highly enhanced coercivity, cluster spin glass, and meta-magnetic behavior. The enhanced coercivity ($\simeq$1~kOe) makes F3GT-2 much more useful for memory storage applications and is likely to elucidate a new route to tune useful magnetic properties. Moreover, this method is much more convenient than hetero-structure and other cumbersome techniques.
\end{abstract}

\maketitle

{\section{Introduction}}
The magnetic materials are an integral part of modern-day technologies, based on desired application-oriented magnetic properties such as saturation magnetization ($M_\text{s}$), remanent magnetization, and coercivity ($H_\text{c}$)\cite{Hirosawa_2017,John_book}. The two-dimensional (2D) van der Waals (vdW) materials have recently emerged as potential candidates  for technological applications \cite{Ding2021, Huang2017, Gong2017, Burch2018}. However,  the typical Curie temperatures and coercitivity of these vdW magnetic materials are low compared to their three-dimensional (3D) counterparts \cite{Kabiraj2020}.

Recently, the family of Fe$_n$GeTe$_2$ ($n$~=~3,~4,~5) (FnGT) vdW compounds \cite{Wang2017b,Mondal2021,BERA2023170257,May2020} are found to be both metallic \cite{Yi2016} and ferromagnetic (FM) with reasonably, higher $T_\text{C}$ ($\gtrsim$ 200 K) compared to other reported families of FM vdW FM materials \cite{Chen2013,Liu2018,Son2019,McGuire2017}. Moreover, this family of vdW FM materials hosts complex magnetic phases with complex spin textures \cite{BERA2023170257} and emerged as versatile scientific platforms for studying various novel quantum phenomena\cite{Geim2013, Geim2007, Novoselov2005}. The Fe$_3$GeTe$_2$ (F3GT) is a well-studied compound of this family, because of the following features - (i) Easy to grow a few millimeter-sized crystals, (ii) Transition temperature, $T_\text{C}$ can easily be tuned in the range, 150-230~K by varying Fe concentration (i.e. $\textit{x}$) in Fe$_{3-x}$GeTe$_2$ ) \cite{Zhu2016}, (iii) Observation of novel phenomena, like heavy-fermion sate\cite{Yun2021}, strong electron correlation effect\cite{Zhu2016} and (iv) manipulable magnetic domain\cite{Nguyen2018}.  These properties make F3GT an important scientific platform  for further research to enhance $T_\text{C}$  and coercivity ($H_\text{c}$) for potential applications.  A recent study revealed that intercalation of F3GT with sodium and TBA+ molecules leads to a ferromagnetic ordering above room temperature (T$>$ 300 K) \cite{Weber2019, Srinivasa2022}. Very recently L. Zhang $et.al.$, have reported an improvement of ferromagnetic transitions temperature, $T_\text{C}$ in the heterostructure of Fe$_3$GeTe$_2$ and FePS$_3$\cite{Zhang2020} due to the proximity coupling effect, which modifies the spin textures of F3GT at the interface. 

However, the low $H_\text{c}$ value of F3GT is a significant hindrance to the realization of prospective technological applications\cite{LeonBrito2016}. The coercivity ($H_\text{c}$) is the measure of the ability of a ferromagnetic material to withstand an external perturbation without becoming demagnetized. Based on the value of coercivity ($H_\text{c}$), magnetic materials are classified into two segments, soft and hard ferromagnets. Soft magnets generally show coercivity ($H_\text{c}$) in the Oe range, whereas later is in the order of kOe range. Soft magnets are required in applications such as switching, sensing, and microwave absorbing medium \cite{Fallarino_2021,LIU201655}, whereas, hard magnets are useful in motors, generators, and data storage devices\cite{Zhu_2021}. Therefore, the search for new high-performance magnetic materials based on transition metals (free from costly rare earth elements), in particular, 2D vdW magnetic materials with high $H_\text{c}$ has been the research focus for the past decade \cite{Zhao2021}. The high $H_\text{c}$ can be tuned or obtained from the combination of several factors, such as intrinsic magnetic interaction \cite{Wang2021,Niu2019}, high pressure\cite{Feng2019}, doping\cite{Ahmed2020,Ahmed2017,Chang2018,Ahmed2020a}, pinning effect by defects or stress\cite{Dho2005,Liu2006,Tan2014}, and shape anisotropy\cite{Zhang2021a,Rout2021}  of nanosheets due to their 2D nature. However, all these methods involve cumbersome synthesis and application procedures and also may alter other important magnetic properties of the sample including $T_\text{C}$, saturation magnetization values, etc., all of which are undesirable from the application point of view. To our knowledge, there are no reports regarding the enhancement of $H_\text{c}$ value in the parent F3GT compound subject to different growth techniques. Tuning growth techniques to synthesize F3GT and subordinate compounds to enhance the required critical coercive field ($H_\text{c}$) can be a promising way for the development of 2D vdW magnetic materials with a high coercive field. Additionally, by tuning of synthesis condition,  one can introduce controlled defects, impose disorder, and change the particle size in the compound, leading to the emergence of many novel magnetic phenomena such as exchange bias effect\cite{Bhanuchandar2019}, glassy, and super-paramagnetic state\cite{Pereira2012} which are not only important for fundamental understanding but also bear technological relevance.\cite{Amit1985,Bryngelson1987a,Pazmandi1999,Guardia2011,Kolhatkar2013}

Here, we demonstrate a very easy, robust, and highly effective method of phase engineering by altering the sample growth conditions to greatly enhance the intrinsic coercivity (7-10 times) of the sample, without compromising its fundamental magnetic properties ($T_\text{C}$$\sim$210~K). Using the method of phase engineering, by altering the sample growth conditions and initial reactant stoichiometry (mentioned in detail in the manuscript), we have successfully been able to enhance the intrinsic coercivity of the sample by an order of magnitude, up to temperatures close to $T_\text{C}$, without any reduction in $T_\text{C}$ and other parameters. The phase-engineered version (F3GT-2) comprises of parent F3GT phase with a small percentage of randomly embedded clusters of a co-planar FeTe (FT) phase, which serve as both mosaic pinning centers between grains of F3GT above the antiferromagnetic transition of FT ($T_\text{C1}$ $\sim$ 70~K) and also as anti-phase domains below $T_\text{C1}$.

A comparative magnetic and magnetotransport study of parent F3GT (F3GT-1) along with the phase-engineered version (F3GT-2) has been performed.
The samples were grown using the chemical vapor transport (CVT) method. Single-crystal x-ray diffraction results confirm F3GT-1 as single-phased, whereas F3GT-2 shows a $1.2\% $ FeTe (FT) phase, which has been deliberately added by altering the growth conditions. The 2-phase structure in F3GT-2 is further verified by electrical transport measurements~\ref{app:Resistivity}. In addition to increased coercivity, other new interesting phenomena, such as cluster glass (CG) and meta-magnetic behavior have also been observed in F3GT-2. The coercive field of F3GT-2 has been successfully enhanced to 7-10 times the value of  $H_\text{c}$ of F3GT-1. This study demonstrates an effective way to significantly enhance the coercivity of 2D VdW ferromagnetic materials in general, for prospective potential applications. 


{\section{Experimental details}}
Two batches of F3GT samples (labeled as F3GT-1 and F3GT-2) were grown in single crystal form by chemical vapor transport (CVT) method with I$_2$ as transport agent in both cases. The high-purity elements: Fe (99.99\,$ \% $ pure), Ge (99.99\,$ \%  $pure) and Te (99.99\,$ \% $ pure) powder were used as starting materials. To grow good quality F3GT crystals (i.e. F3GT-1), the powders were mixed stoichiometrically (molar ratio, Fe : Ge: Te = 3 : 1: 2) and grind thoroughly in an agate motor. Then 5 mg-cm$^{-3}$ I$_2$ powder was added to the mixture and placed in a quartz tube. The tube was sealed under a vacuum of 2$\times 10^{-4}$ mb,  thereafter the sealed quartz tube was placed horizontally in a two-zone horizontal tube furnace for seven days, keeping the hot-end at 750$^\circ$C and cold-end at 700$^\circ$C, respectively. F3GT-2 crystals were also prepared by the same method with different temperature gradients (cold zone at 700$^\circ$C and hot zone at 800$^\circ$C). Shining single crystals with dimensions $2\times1.5\times0.1~mm^3$ and $2.5\times2.5\times0.5~mm^3$ were obtained at the cold-end for both the cases respectively\cite{Sultan2021,Chen2013}. The phase composition and crystal structure were investigated by the standard X-ray diffraction (XRD) technique using Rigaku diffractometer equipped with Cu-$K_\alpha$radiation ($\lambda$= 1.54056 \AA). The micro-morphology was characterized with a field emission scanning electron microscope (FE SEM, JEOL LSM-6500). The elemental composition analysis of both samples was done by taking x-ray energy-dispersive spectroscopy (EDS).  Electrical transport measurements were carried out by an Oxford Instrument cryogenic system with a conventional dc measurement method, from 5~K - 300~K. Magnetic studies (\textit{M-T}, \textit{M(H)}, and \textit{M-t}) were done using Magnetic Properties Measurement System (MPMS-SQUID) in the temperature interval of 2~K-300~K, up to 50 kOe magnetic field \textit{H}. The plate-like single crystals were used to minimize the misalignment of samples during these measurements. 

\begin{figure}
	\centering
	\includegraphics[width=1.0\columnwidth]{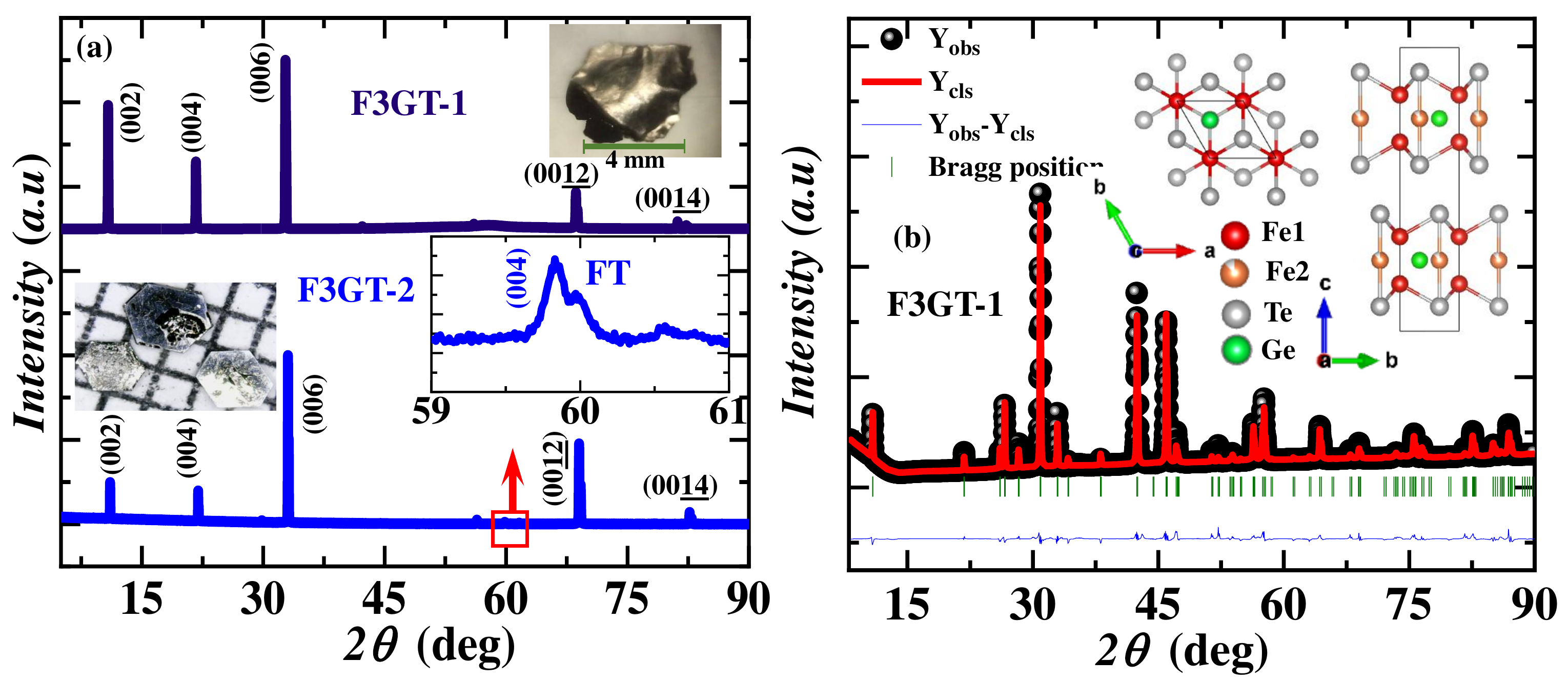}
		\caption{\textbf{Structural analysis of F3GT crystal.}  (a)~The X-ray diffraction pattern obtained from the cleaved plane of F3GT (F3GT-1 $\&$ F3GT-2) single crystals at room temperature. The picture of the single crystal is shown in the inset. (b) Observed (solid black sphere) and calculated (solid red line) powder X-ray-diffraction patterns for the powder F3GT-1  crystals. The green short lines denote the Bragg positions, and the blue curve indicates the difference between the observed and calculated patterns. View of the crystallographic structure of F3GT-1 from the \textit{a} axis and \textit{c} axis. The ash,  red, and green spheres represent the Te, Fe and Ge respectively. The black box with the cross-section rectangles depicts a crystallographic unit cell.}
		\label{fig:XRD}
\end{figure}

{\section{Results}}

\subsection{XRD analysis and crystal structure}

The room temperature single-crystal XRD patterns of F3GT-1 and F3GT-2 are shown in Figure~\textcolor{blue}{\ref{fig:XRD}(a)}. Powder XRD data of the F3GT-1 sample are well fitted the P63/mmc space group and lattice parameters are $a$ = 3.96 $\text{\AA}$ and $c$ = 16.36 $\text{\AA}$ (detail analysis are presented in  Figure~\textcolor{blue}{\ref{fig:XRD}(b)}). In the single-crystal XRD pattern, all peaks come from (00\textit{l}), which indicates that the crystal surface is normal to the \textit{c }axis with the plate-shaped surface parallel to the ab plane. Besides, the most important point is that two small peaks are observed for F3GT-2 (see Figure~\textcolor{blue}{\ref{fig:XRD}(a)}) around 2$\theta$ = 60$^\circ$. These secondary peaks indicate the existence of the   FeTe phase\cite{Junho2021}. By comparing with the profile peak the obtained percentage of the secondary phase is 1.22 \%.
The crystal structure of F3GT-1 is shown inset of the Figure~\textcolor{blue}{\ref{fig:XRD}(b)}, which contains Fe$_3$Ge slabs isolated by van der Waals bonded Te double layers. The Fe atoms in the unit cell contain two non-identical Wyckoff sites Fe1 and Fe2, as shown in the inset of the Figure~\textcolor{blue}{\ref{fig:XRD}(b)}. The Fe1 atoms are situated in a hexagonal net in a layer with only Fe atoms. The Fe2 and Ge atoms are covalently bonded in an adjacent layer\cite{Wang2017b, Mao2018}. By contrast, our EDX result gives Fe deficiencies and Te excess for F3GT-1 with a composition of Fe$_{2.84}$GeTe$_{2.1}$ and for F3GT-2 with a composition of Fe$_{2.94}$GeTe$_{2.12}$.

\begin{table}
	\caption{Crystallographic parameters obtained from a Rietveld refinement of the powder x-ray diffraction pattern collected at room tempareture for a polycrystalline sample with a nominal composition Fe$_3$GeTe$_2$(F3GT-1) } 
	\centering 
	
	\begin{tabular}{|p{4cm}|p{4cm}|}
		\hline
		Nominal composition & Fe$_3$GeTe$_2$  \\
		Refined composition & Fe$_{2.88}$GeTe$_2$  \\
		Structure & Hexagonal\\
		Space group & \textit{P}6$_3$/\textit{mmc} (No. 194) \\
		Formula units/unit cell (Z) & 2 \\
		Lattice parameters &  \\
		a(\AA) & 3.96148 \\
		b(\AA) & 3.96148 \\
		c(\AA) & 16.3812 \\
		V$_{cell}$(\AA$^3$) & 222.6339 \\
		\hline
	\end{tabular}\\
	\begin{tabular}{|p{1.0cm}|p{1.33cm}|p{1.2cm}|p{1.2cm}|p{1.2cm}|p{1.5cm}|}
		\hline
		Atom & Wyckoff position & x & y & z & Occupancy \\
		\hline
		Fe1 & 4e & 0 & 0 &  2/3 & 1 \\
		Fe2 & 2c & 2/3 & 1/3 & 3/4 & 0.882 \\
		Ge1 & 2d & 1/3 & 2/3 & 3/4 & 1 \\
		Te1 & 4f & 2/3 & 1/3 & 0.58709 & 1 \\
		\hline
	\end{tabular}
\end{table}

\subsection{Magnetic properties of F3GT:}

\subsubsection{Magnetization vs temperature measurements}
\begin{figure}
		\centering
		\includegraphics[width=1\columnwidth]{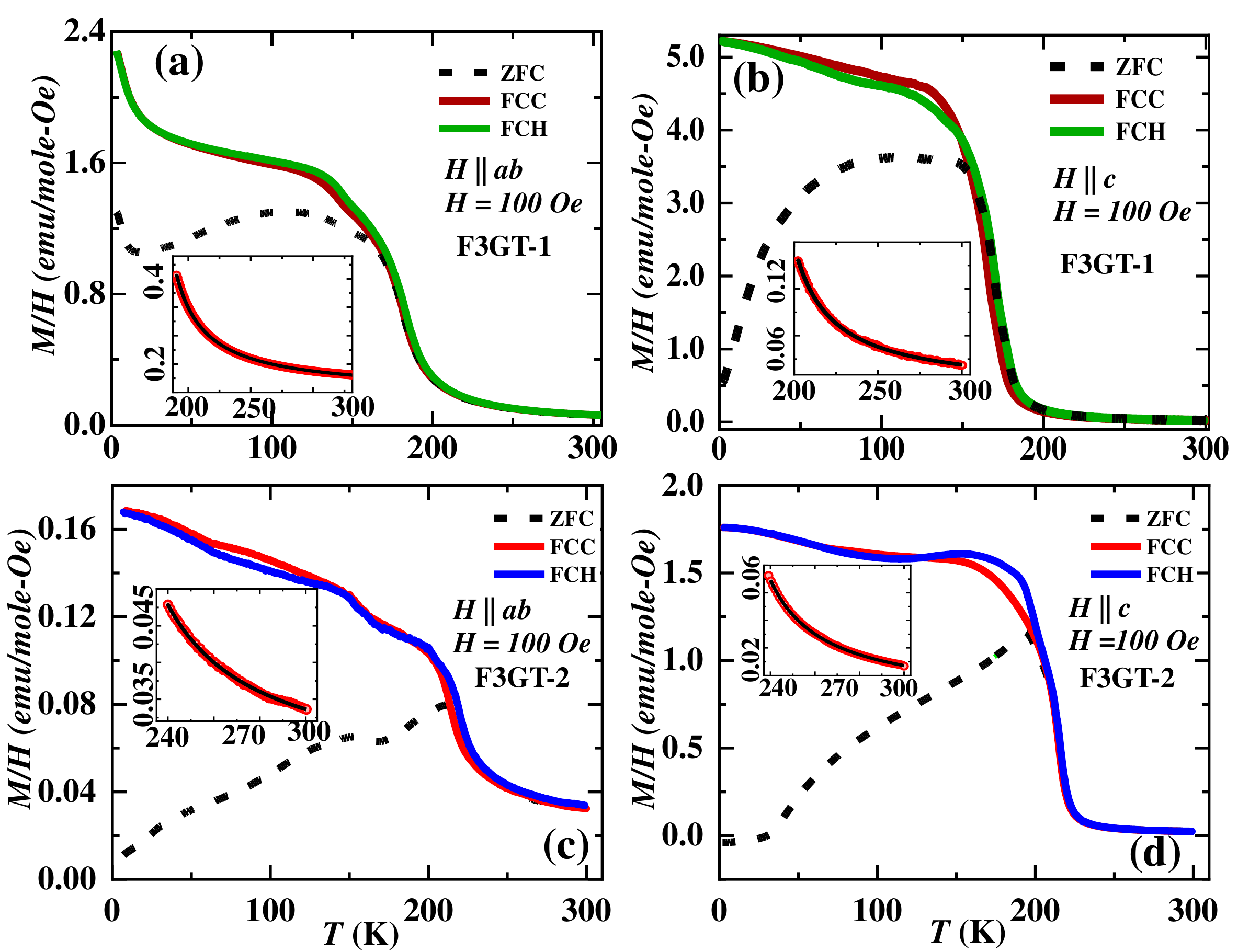}
		\caption{\textbf{Temperature dependent magnetization measurements.} Magnetization measured at external magnetic field \textit{H }= 0.1 kOe applied in the \textit{ab} plane and along the \textit{c} axis under zero-field-cooling (ZFC), field-cooling (FC) and field-cool-heating (FCH) protocols for (a-b) F3GT-1 and (c-d) F3GT-2. The inset shows modified Curie-Weiss fits to the magnetic susceptibility data. }
		\label{fig:MT}
	\end{figure}

The dc magnetic susceptibility ($\chi_{dc}$= \textit{M/H}) of both samples have been measured in the zero-field-cooled (ZFC), field cool cooling (FCC), and field-cool-heating (FCH) protocols under applied magnetic fields (H) in the temperature range 2 to 300~K. Figure~\textcolor{blue}{\ref{fig:MT}(a-b)} shows the \textit{M-T} data of F3GT-1 at applied magnetic field, \textit{H} = 100 Oe in \textit{ab} plane and parallel to the \textit{c} axis, respectively.   Initially, the susceptibility increases slowly with decreasing temperature demonstrating a typical paramagnetic (PM) nature. With a further decrease in temperature, the susceptibility rapidly increases at around \textit{T$_{C}$}, revealing the onset of a ferromagnetic (FM) ordering\cite{Wang2017b, Mao2018, Chen2013}. The effective magnetic moment of F3GT-1 is calculated from by fitting the $M/H$(T) data (using\textit{H} = 0.1 kOe FCC data) using following modified \textit{Curie-Weiss} (C-W) law,

	\begin{center}
		\begin{equation}
			\begin{split}
				\frac{M}{H}=\chi_0+\frac{C}{(T-\Theta_p)}  \\
			\end{split}
		\label{eq:CW_law}
		\end{equation}
	\end{center}
	
in the temperature range 240-300 K, where C is the Curie constant and $\Theta_p$ is the C-W temperature (see inset of Figure \textcolor{blue}{\ref{fig:MT}(a)}). The obtained effective magnetic moment $(\mu_{eff})$ and C-W temperature ($\Theta_p$) from the fit are 4.54 $\mu_B$/f.u. and 162 K for $H\parallel ab$, respectively. Similarly, we have obtained $(\mu_{eff})$  = 4.61 $\mu_B$/f.u. and  $\Theta_p$ = 178 K for $H\parallel c$ direction(see inset of Figure \textcolor{blue}{\ref{fig:MT}(b)}). The positive value of $\Theta_p$ indicates the prominent FM interaction of the compound. These effective moment values are in good agreement with earlier reports\cite{Wang2017b, Liu2017}.

The first derivative of $\chi_{dc}$ w.r.t  \textit{T} ($d\chi_{dc}/dT$) data of F3GT-1 along the $ab$ plan (see the Figure~\textcolor{blue}{\ref{app:Magnetization}} (a-b) in appendix) reveals another two magnetic ordering at around \textit{T$_{C1}$} = 142 K and \textit{T$_{C2}$} = 6 K. The above signatures in susceptibility data strongly suggest complex nature of the magnetic ground state in F3GT with novel magnetic properties \cite{BERA2023170257}.

Figure~\textcolor{blue}{\ref{fig:MT}(c-d)} presents the \textit{$\chi_{dc}$(M/H)-T} nature of F3GT-2 at H = 0.1 kOe, in  \textit{ab} plane and parallel to the \textit{c} axis, respectively and shows a ferromagnetic ordering at \textit{$T_\text{C}$}. The $d\chi_{dc}/dT$ vs \textit{T} for F3GT-2 suggests that besides FM transition, there exists another magnetic ordering at around 70 K (see appendix \textcolor{blue}{\ref{app:Magnetization}}) confirming the presence of FT phase as observed in XRD study \cite{Junho2021,Jiang2013}. The ZFC and FCC plots show significant splitting below 210 K for $ H\parallel ab$ and $H \parallel c$ directions. The extent of bifurcation between FCC and ZFC data also increases with decreasing values of H (for details see appendix \textcolor{blue}{\ref{app:Magnetization}}). Such behavior is often observed in disordered or glassy magnetic systems. Below $T_\text{C}$ = 210 K, the value of $\chi_{dc}$ for $H\parallel c$ axis is about eleven times larger than it for $H\parallel ab$ plane, which indicates the anisotropic behavior in the ordered state \cite{Mao2018, Liu2018b}. 
The high-temperature (240-300 K) \textit{M/H} data is fitted (see the inset of Figure~\textcolor{blue}{\ref{fig:MT}(c-d)}) using modified Curie-Weiss law (equation~\ref{eq:CW_law}). The obtained paramagnetic moment is $\mu_{eff}$ to be 2.94$\mu_B$/(f.u. of Fe$_{2.94}$GeTe$_{2.1}$ ), and $\Theta_p$ =208K for $H \parallel c$ and 198K for $H \parallel ab$. The value of $\chi_0$ is found to be small but positive\cite{Wang2017b, Liu2018b}.
 
\subsubsection{Magnetization vs applied magnetic field measurements: M-H loops}

\begin{figure}[ht]
		\includegraphics[width=1\columnwidth]{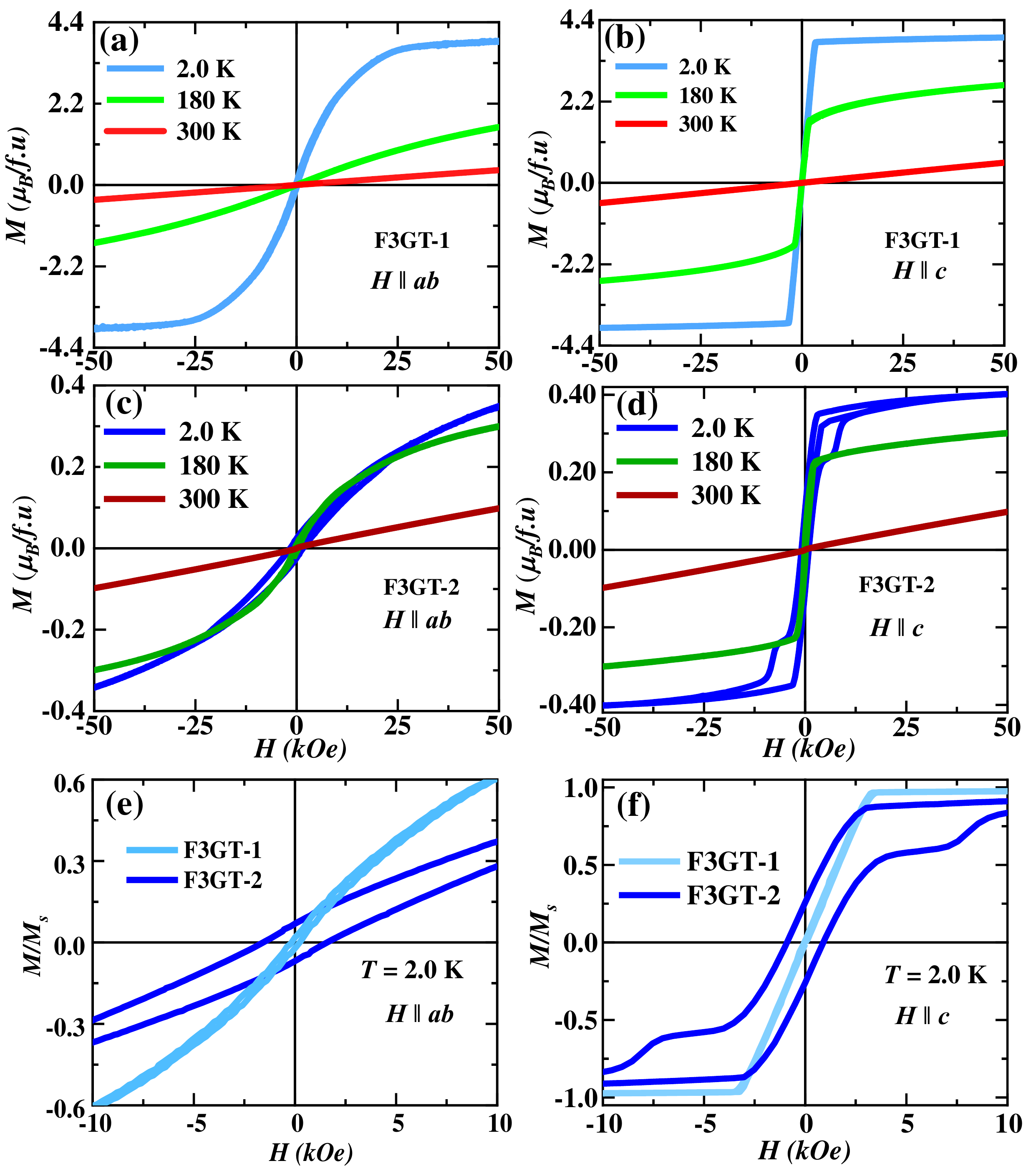}
		\caption{\textbf{Isothermal magnetization measurements.} Isothermal magnetization at few selected temperatures measured in the \textit{ab} plane and along the \textit{c} axis for (a-b) F3GT-1 and (c-d) F3GT-2. (e-f) Normalized \textit{M/M$_s$} at 2K for both samples shows significant enhancement of $H_{C}$ and the emergence of metamagnetic feature in F3GT-2 with reference to F3GT-1.}
		\label{fig:MH}
	\end{figure}

To shed light on the nature of the magnetic phase, measurement of magnetization as a function applied magnetic field was carried out at various temperatures, as shown in Figure~\textcolor{blue}{\ref{fig:MH}}. The isothermal $M$ vs $H$ curves of F3GT-1 sample at a few selected temperatures(\textit{T}=2,~180 and 300 K) in \textit{ab} plane and along \textit{c}-axis are presented in Figure~\textcolor{blue}{\ref{fig:MH}(a and b)} respectively. The saturation behavior is observed in both directions. The saturation fields at 2 K are $H_c^{s}\approx$ 3.34 kOe for $ H \parallel c$ axis and $H_{ab}^{s}\approx$ 27 kOe for $H \parallel ab$ plan which suggests that the easy magnetization direction is along the \textit{c} axis. Moreover, the saturation magnetic moments at 2~K is M$_c^{s}$ = 4.01$\mu_B$/formula for $H \parallel c$ axis, and M$_{ab}^{s}$ = 3.38$\mu_B$/formula for $H \parallel ab$ plane. These results are in good agreement with the previous report\cite{Wang2017b, Liu2018b}. The saturation magnetization for in-plane direction is lower than the out of plane direction which suggests the existence of a substantial amount of the AFM phase coexisting along with the dominant FM phase in the F3GT-1 sample\cite{Fita2003}.\\

Figure~\textcolor{blue}{\ref{fig:MH}(c and d)} shows the \textit{M(H)} curves of F3GT-2  in both directions at a few selected temperatures. The saturation applied field at \textit{T}~=~2 K, $H_s^c\approx$ 4kOe for $H \parallel c$ is much smaller than $H_s^{ab}\approx$ 40kOe for $H \parallel ab$, confirming the easy axis is the \textit{c} axis. The saturation moment at \textit{T}~=~2~K is $M_s^c$= 0.41$\mu_B$/f.u for $H \parallel c$ and $M_s^{ab}$~=~0.38$\mu_B$/f.u for $H \parallel ab$ , respectively. Notably, F3GT-2 sample shows a significantly large coercive field of $H_c^c$$\approx$ 950 Oe for $H \parallel c$ and $H_{c}^{ab}$$\approx$1550 Oe for $H \parallel ab$ at 2 K, 
which are higher than the F3GT-1\cite{Liu2018b}. The sudden slope change after reaching a certain critical field (15 kOe) a step-like feature appears that relates to the meta-magnetic transition \cite{Bennett2016,Murthy2014}. The obtained $H_\text{c}$ for F3GT-2 is significantly larger than F3GT-1 at~\textit{T}~=~2 K (almost 12 times).  Whereas the saturation magnetization $M_s^c$ is 0.41$\mu_B$/f.u for $H \parallel c$, which is about $\sim$~10 order of magnitude lower than the F3GT-1. The value of $M_s$ and $H_\text{c}$ in both the directions for F3GT-1 and F3GT-2 are presented in tabular form in Table~\ref{table:MH}.
	
	 \begin{table}[ht]
		\caption{Saturation magnetization (M$_s$) and coercivity(H$_c$) of F3GT  at T$\sim$2 K. Shows significant enhancement of coercivity in F3GT-2.} 
		\centering 
		\begin{tabular}{ c c c c} 
			\hline 
			Sample ~~& Direction ~~ & ~~M$_s$($\mu_B/f.u$)~~ &~~ H$_c$(Oe) \\ [0.50 ex] 
			\hline 
			F3GT-1 &  H$\parallel$c & 4.01 & 45 	 \\ 
			& H$\parallel$ab& 3.38 & 230 \\
			\hline
			F3GT-2&  H$\parallel$c & 0.41 & 950 \\
			& H$\parallel$ab& 0.38 & 1550\\
			\hline 
		\end{tabular}
		\label{table:MH} 
	\end{table}
The Rhodes-Wohlfarth ratio (RWR) is a very good criterion to ascribe whether the magnetic material belongs to the itinerant or a localized spin system. In the Stoner model, RWR is defined as $\mu_c/\mu_s$, where $\mu_c(\mu_c$ + 2) = $\mu^2_{eff}$ and $\mu_s$ is the spontaneous magnetization in the ground state. RWR is 1 for a localized system and greater than 1 for an itinerant system. The RWR of F3GT-1 is found to be 3.25, while RWR of F3GT-2 is 5.52, suggesting itinerant character and/or strong spin fluctuations in the ground state\cite{Liu2017}.\\

\begin{figure}
		\includegraphics[width=1\columnwidth]{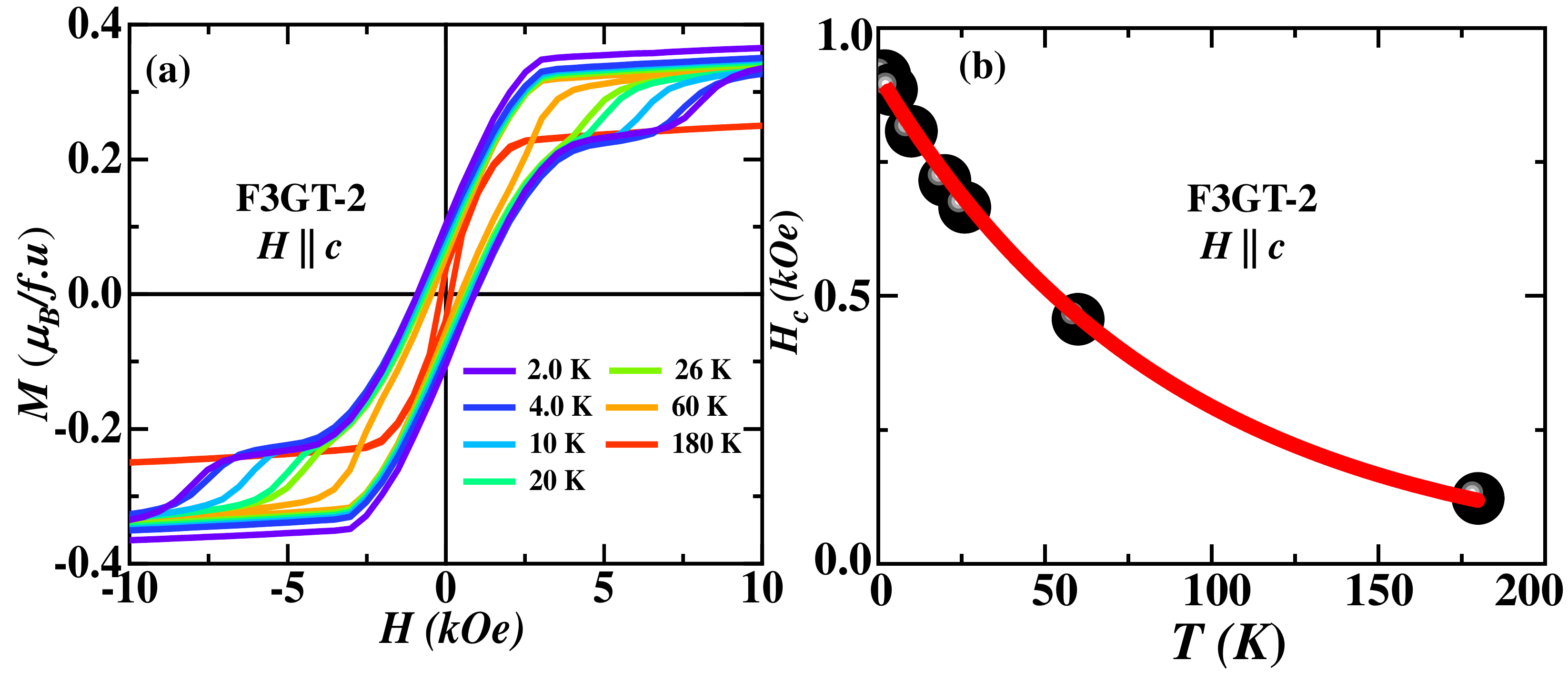}
		\caption{\textbf{Determination of coercive field, $H_\text{c}$ of F3GT-2 from isothermal magnetization.} (a) Isothermal magnetization  measured at various temperatures (\textit{T}=2, 4, 10, 20, 26, 60 and 180 K) between $\pm$ 50 kOe for F3GT-2 ($ H\parallel c$). (b) Coercive field ($H_\text{c}$) vs.temperature ($T$) plot and fits using equation~\ref{eq:expo} (red solid line) expected for spin glass/cluster spin glass systems.}
		\label{fig:MHT}
\end{figure}

Narrow\textit{ M(H)} hysteresis loop with non-zero remanence and coercivity without saturation below $T_\text{C}$ has been reported in spin-glasses/cluster spin glasses  as well in superparamagnetic (SPM)  systems \cite{Dong2011,Vafaee2016,Sharma2018a}. A distinction between the \textit{M(H)} loops of spin-glasses/cluster spin-glass and SPM systems can be made through the study of the temperature variation of coercivity. Therefore, \textit{ M(H)} hysteresis loops are recorded at various temperatures and presented in Figure~\textcolor{blue}{\ref{fig:MHT}(a)}. 
It is seen that temperature dependence of coercive field $(H_c)$ determined from \textit{M(H)} loops is shown in Figure~\textcolor{blue}{\ref{fig:MHT}(b)}. The coercivity  exponentially decreases with increasing temperature and follows the empirical relationship expected for spin glasses/cluster spin glasses below the transition temperature $T_\text{C}$ $\sim$210 K\cite{Kumar2020a, Sharma2018a}

	\begin{center}
		\begin{equation}
			\begin{split}
				H_c(T)=H_c(0)~e^{-\alpha T}  \\
			\end{split}
			\label{eq:expo}
		\end{equation}
	\end{center}
 
 where $H_\text{c}$(0) is the coercive field at 0 K, while $\alpha$ is a fitting parameter. The solid line in Figure~\textcolor{blue}{\ref{fig:MHT}(b)} is the least squares fit using equation~\ref{eq:expo} with fitting parameters $H_\text{c}$(0)=(911 $\pm$ 14)~Oe, $\alpha$= (0.0112 $\pm$0.0006)~K$^{-1}$. The exponential dependence of $H_\text{c}$ is regarded as the signature of spin glasses/cluster spin glasses\cite{Sharma2018a}.

\subsubsection{Magnetic relaxation $\&$ ac susceptibility:}
\begin{figure}
		\includegraphics[width=1\columnwidth]{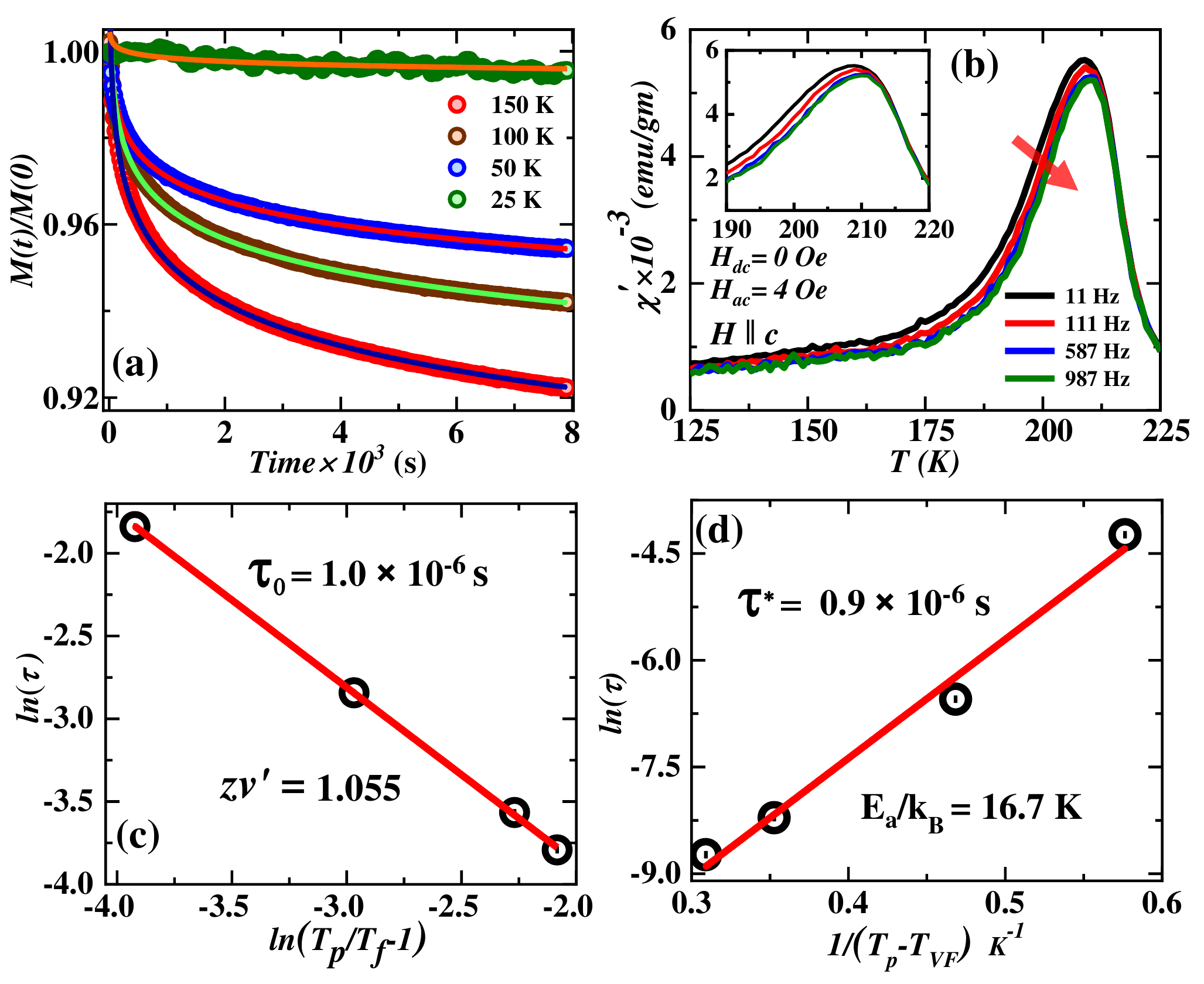}
		\caption{\textbf{Magnetic relaxation and the ac susceptibility ($\chi_{ac}^\prime$(\textit{T})) }(a) Relaxation of \textit{M} at different temperatures after the field is withdrawn. The solid lines are fits with a logarithmic relaxation function (see equation~\ref{eq:relaxation}). (b) The real parts of the ac magnetic susceptibility ($\chi_{ac}^\prime$(T)) of F3GT-2 measured at a few selected frequencies from 11 Hz to 987 Hz in an applied ac magnetic field of 4 Oe along \textit{c}-axis. (c) The relaxation time ($\tau$) dependence of freezing temperature plotted as ln($\tau$) vs ln(\textit{t}). Here  \textit{t} = (T$_p$ - T$_{f}$)/T${f}$ is reduced temperature. The solid line represents the fit to the power-law divergence. (d) Relaxation time ($\tau$) vs freezing temperature plotted in log scale ( ln($\tau$) vs 1/(T$_p$ - T$_{VF}$)) together with the fit using Vogel-Fulcher equation.}
			\label{fig:AC_chi}
\end{figure}

The above $M$($T$) and $M$($H$) results
 directly point towards the presence of disorder and/or frustrations in F3GT-2. To probe the stability of the zero-field magnetic state, magnetic relaxation at a few selected temperatures (\textit{T} = 25, 50, 100, and 150 K) are measured in the FCC condition and presented in Figure~\textcolor{blue}{\ref{fig:AC_chi}(a)} \cite{Singh2010}. The sample is first cooled from 300 K to the desired temperature in the presence of \textit{H} = 100 Oe ($H \parallel c$) and \textit{M} is measured as a function of time(t) immediately after removing the field. The magnitude of relaxation is found to increase with increasing \textit{T}. After \textit{t} = 2 h, the sample shows about 10$\%$ change in \textit{M} at 150 K, which is enormous as far as the relaxation in other glassy and disordered systems are concerned \cite{Chatterjee2020}. This relaxation behavior can be fitted with a logarithmic relaxation function given by,
 
	\begin{center}
		\begin{equation}
			\begin{split}
				M(t)=M_0+\alpha~\log(t)  \\
			\end{split}
		\label{eq:relaxation}
		\end{equation}
	\end{center}

here \textit{M$_0$} is the initial magnetization and $\alpha$ is the relaxation rate in dynamic equilibrium\cite{Singh2010, Anand2012a}.\\

Although the logarithmic time dependence of isothermal remanent magnetization is normally associated with glassy magnetic system, it has also been observed in materials with complex interactions without spin-glass freezing. Therefore, to further check whether the relaxation is related to the formation of a glassy state,the temperature dependent AC susceptibility ($\chi_{ac}$(\textit{T})) measurement was performed in temperature regime from 240K to 75K at various frequencies (11 Hz, 111Hz, 587 Hz and 987 Hz ) under an AC drive field of 4 Oe \cite{Singh2010}. The real part of susceptibility,  $\chi_{ac}^\prime$(\textit{T}) exhibits anomaly . The amplitude and peak position depend on the frequency of the applied ac magnetic field. The maxima in  $\chi_{ac}^\prime$(\textit{T}) observes around 208.9 K at 11 Hz, shown in Figure~\textcolor{blue}{\ref{fig:AC_chi}(b)}, support the $T_\text{C}$ peak in the dc magnetization curves. The position of maxima  shifts towards the higher temperature as the frequency is increased, which further confirms the existence of a glassy magnetic phase. It is also observed that the amplitude of maxima in $\chi_{ac}^\prime$(T) decreases with increasing frequency. The Mydosh parameter is used as the criterion to categorize the glassy magnetic systems according to the response of magnetic spins compare the relative shift in freezing temperature per decade of frequency\cite{Mulder1981}. To understand the spin freezing dynamics, the Mydosh parameter is calculated followed by Vogel-Fulcher (VF) and critical slowing dynamics models \cite{Yadav2019}. The Mydosh parameter (K) is defined as
\begin{center}
	\begin{equation}
		\begin{split}
			K=\frac{\bigtriangleup T_{pf}}{T_p\bigtriangleup(log(f))}  \\
		\end{split}
	\end{equation}
	\label{eq:Mydosh}
\end{center}
where $T_p$ represents the temperature where the AC susceptibility is maximum and f is the measured frequency. Here,  $\bigtriangleup T_{pf}$ = $T_{Pf1}$-$T_{Pf2}$ and $\bigtriangleup log(f) = log(f1)-log(f2)$ .  The value of K is 0.004 which is comparable to the reported values for other Cluster Glass (CG) systems (\textit{K} $\leq$0.08) \cite{Das2018,V.K.2020a}{\normalsize }.\\

The frequency dependence of $T_p$ follows the conventional power-law divergence of critical slowing down,
	\begin{center}
		\begin{equation}
			\begin{split}
				\tau=\tau_0 (\frac{T_p-T_{f}}{T_f})^{-{z\nu^\prime}}  \\
			\end{split}
		\label{eq:powerlaw}
		\end{equation}
	\end{center}

where $\tau$ is the relaxation time corresponding to the measured frequency($\tau=1/\nu$), $\tau_0$ is the characteristic relaxation time of the individual spin cluster, $T_f$ is the freezing temperature for $\nu\rightarrow$0 Hz, and ${z\nu^\prime}$ is the dynamic critical exponent[$\nu^\prime$ is the critical exponent of correlation length, $\xi= (T_p /T_f -1)^{-\nu^\prime}$ and the dynamical scaling relates $\tau$ to $\xi$ as $\tau\thicksim\xi^z$]\cite{Mukadam2005, Anand2012,Mydosh2015}.

	It is useful to rewrite equation~\ref{eq:powerlaw} as
	\begin{center}
		\begin{equation}
			\begin{split}
				ln(\tau)=ln(\tau_0)- {z\nu^\prime}ln(t)  \\
			\end{split}
		\label{eq:logpow}
		\end{equation}
	\end{center}

	where t=$\frac{T_p-T_{f}}{T_f}$. The plot of ln($\tau$) versus ln(t)is shown in the Figure~\textcolor{blue}{\ref{fig:AC_chi}(c)}, from the slope and intercept we can estimate the value of ${z\nu^\prime}$ and $\tau_0$ respectively. Obtained values of the characteristic relaxation $\tau_0$ , ${z\nu^\prime}$ and $T_f$ are $1.0\times 10^{-6}$ s, 1.05$\pm$0.05 and 208.88 K for this system\cite{V.K.2020a}. However, the value of $\tau_0$ is larger compared to the typical canonical spin glass systems ( $\sim10^{-12}$ s for). This suggests the possibility of strongly interacting clusters rather than individual spins\cite{Anand2012} in F3GT-2, which gives a slow spin relaxation.

Further, the Vogel-Fulcher (VF) relaxation model has been used to analyze the behavior of interaction among the frozen spin clusters \cite{Anand2012}. The interacting particles (or spin clusters) in a a magnetic system undergoes a relaxation process in which the relaxation time follows the VF law given by,

	\begin{center}
		\begin{equation}
			\begin{split}
				\tau=\tau^\ast exp(\frac{-E_a}{k_B(T_p-T_{VF})})  \\
			\end{split}
		\end{equation}
	\end{center}
	
where $\tau$ is the relaxation time, $E_a $ is the activation energy, $\tau^\ast$ is the characteristic time between the relaxation attempts of spin clusters, and $T_{VF}$ is the VF temperature,  which provides inter-clusters interaction strength. We calculate the value of Vogel-Fulcher temperature $T_{VF}$ by Souletie and Tholence method and obtain $T_{VF}$=207.16 K, which we use to find  $\tau^\ast$ and $E_a $. The linear fitted ln($\tau$)versus $1/(T_P-T_{VF})$ curve is shown in the Figure~\textcolor{blue}{\ref{fig:AC_chi}(d)}. From the linear fitting parameter, we get  $E_a/k_B$=16.7K and $\tau^\ast$=0.9$\times 10^{-6}$, which falls within the range of characteristic relaxation times for the cluster glass system. Thus, critical slowing down of dynamics and VF models along with the Mydosh parameter strongly support spin cluster glass behavior in F3GT-2. 


\section{Discussions}
The structural analysis confirms the single-phase crystalline nature of F3GT-1, whereas it reveals the presence of a
minor FT phase ($\lesssim$ 1.2$\%$)  in the F3GT-2 crystal. Presence of minor FT phase is further confirmed by the resistivity
measurements, where a hump in resistivity at around 68 K (close to the reported $T_\text{C1}$$\sim$~70~K, see appendix~\ref{app:Magnetization}) is observed. The appearance of MI transition is direct evidence, as FT shows a similar nature at $\sim$ 70~K \cite{Kang2020,Maheshwari2015} and ordered antiferromagnetically at around 70~K\cite{Kim2019}. 

		\begin{figure}
		\includegraphics[width=1.0\linewidth]{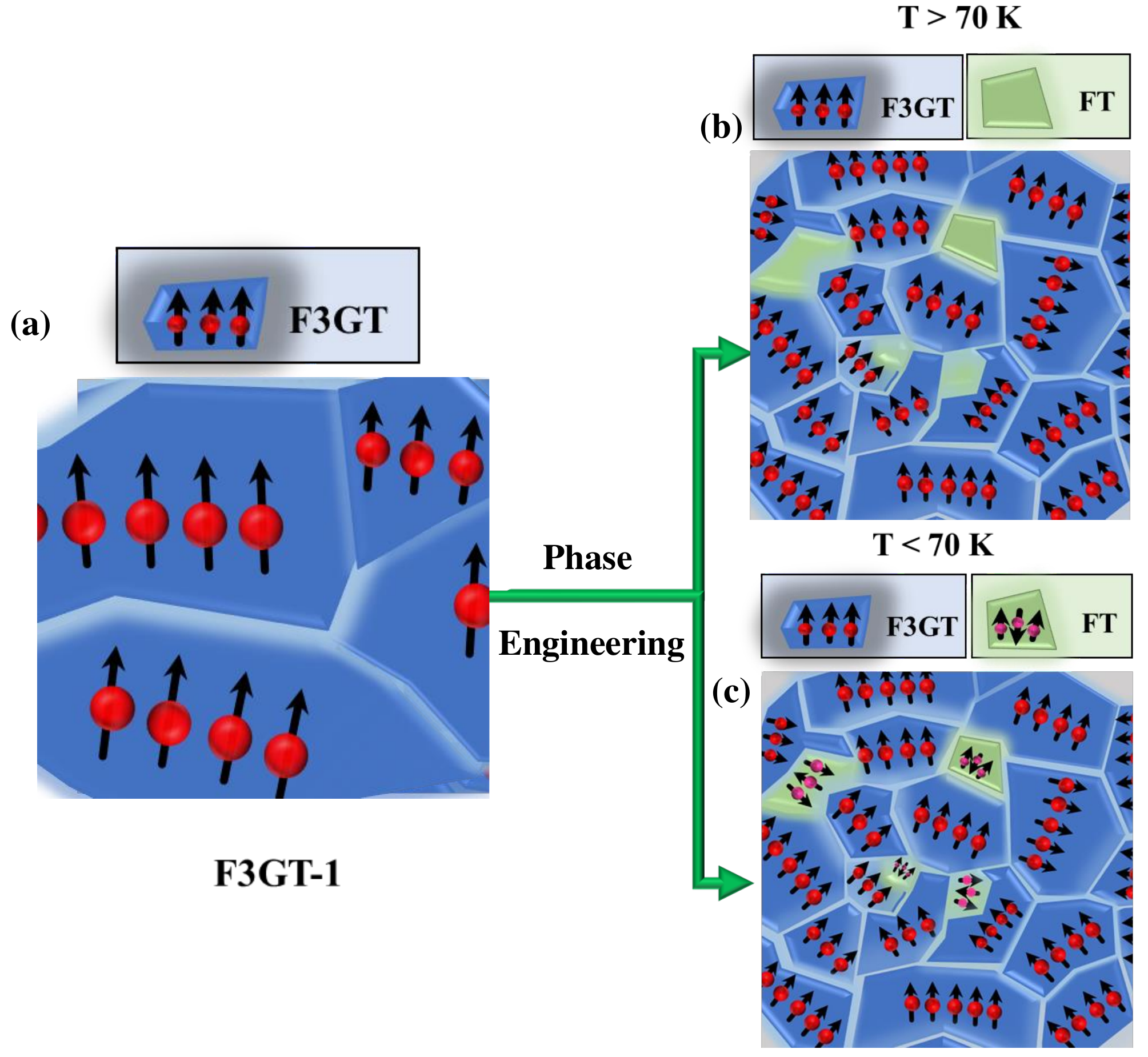} 
		\caption{\textbf{Illustration of conventional cluster spin-glass in F3GT-2.} The schematic diagram of (a) the pure F3GT-1 phase with a relatively smaller amount of multi-domain boundary and (b-c) F3GT-2 sample with a much higher amount of multi-domain boundary disorder above and below ~70~K.}
		\label{fig:schematic}
	\end{figure}
Now based on all the above observations, a schematic micro-structure diagram describing the comparative scenarios of the two samples F3GT-1 and F3GT-2 has been proposed in Figure~\textcolor{blue}{\ref{fig:schematic}}, where the FM and AFM phase domains are presented by light blue and green color respectively and arrow denotes the spin directions. The figure~\textcolor{blue}{\ref{fig:schematic}~(a)}  describes the pure F3GT-1 phase with a relatively smaller amount of multi-domain boundary disorder (small number of magnetic domains with different spin orientations), giving rise to relatively lower spin glass temperature (manifested by the onset of bifurcation in magnetic susceptibility at 182~K in figure~\textcolor{blue}{\ref{fig:MT} (a) } and figure~\textcolor{blue}{\ref{fig:dM_dT} (a-c)} in appendix~\textcolor{blue}{\ref{app:Magnetization}}).

The figure~\textcolor{blue}{\ref{fig:schematic}(b-c)} describes the F3GT-2 sample with a much higher amount of multi-domain boundary disorder (larger number of magnetic domains in the figure), giving rise to much higher spin glass temperature (manifested by the onset of bifurcation in magnetic susceptibility at 215~K, figure~\textcolor{blue}{\ref{fig:MT}~(c)} of \textcolor{blue}{\ref{app:Magnetization}} ). This explains the fundamental change induced by phase engineering, i.e., the increase in disorder and multi-domain formation due to the randomly embedded clusters of the co-planar FeTe (FT) phase, which serve as mosaic pinning centers between grains of F3GT. The presence of FT in F3GT acts as phase separation disorder that separates the FM domains of F3GT. The presence of disorder in the form of mosaic FT phase regions reduces the interdomain magnetic coupling and enhances the spin cluster glass behavior in F3GT-2 \cite{Anand2021a,Blasco2016,Liu2020}. This picture is also consistent with the SEM data, which shows relatively smooth terrain in F3GT-1 but enhanced granularity and the presence of FT phase hexagonal flakes in F3GT-2 as seen in figure~\ref{fig:SEM}.

The other comparative picture (figure~\textcolor{blue}{\ref{fig:schematic}(b-c)}  of F3GT-2) relates to the spin scenario of the same sample above and below $T_\text{C1}$ ($\sim$70K). Above $T_\text{C1}$ the contribution of the FT phase domains is purely structural and non-magnetic disorder related. But below $T_\text{C1}$, the AFM spin texture also starts to contribute to the scenario, giving rise to the metamagnetic features observed in Fig~\textcolor{blue}{\ref{fig:MH}} and figure~\ref{fig:MHT}. The strong anisotropic pinning forces across the anti-phase boundaries (APBs) of the anti-parallel domains offer a significant hindrance to the applied magnetic field (H) to orient them along the field direction \cite{Anand2021a,Blasco2016,Liu2020}. On reaching a critical value of applied H, the magnetization suddenly increases in the form of steps and gives rise to a meta-magnetic transition.

\section{Conclusion}
We have demonstrated an effective way to enhance the coercivity in 2D vdW magnetic material, F3GT through phase engineering, in which the F3GT crystal has been infused with a very small amount of FeTe phase ($\lesssim$ 1.2$\%$). This FT phase acts as both mosaic pinning centers between grains of F3GT above its antiferromagnetic transition temperature ($T_c1\sim$70~K) and also as anti-phase domains below $T_\text{C1}$. This results in increased grain boundary disorder and metastable nature, which leads to highly enhanced coercivity ($\simeq$1~kOe), cluster spin glass, and metamagnetic behavior. The enhanced coercivity makes F3GT-2 much more useful for memory storage applications and is likely to open a new direction toward the development of magnetic materials with higher coercivity and novel magnetic states for potential applications.	
		
\section{Acknowledgement}
This work was supported by the (i) 'Department of Science and Technology', Government of India (Grant No. SRG/2019/000674 and EMR/2016/005437), and (ii) Collaborative Research Scheme (CRS) project proposal(2021/CRS/59/58/760). S.Bera $\&$ A.Bera thanks CSIR Govt. of India for Research Fellowship with Grant No. 09/080(1110)/2019-EMR-I  $\&$ 09/080(1109)/2019-EMR-I, respectively. S.B. acknowledges the experimental facilities for sample growth, procured using financial support from DST-SERB grant nos. ECR/2017/002 037. The authors also would like to acknowledge Mr. Md Yousuf Sk, Prof. Subham Majumdar, and Dr. Subhadeep Datta for technical help and fruitful discussions.	

	\appendix
	\renewcommand{\thefigure}{A\arabic{figure}}	
	\setcounter{figure}{0}	
 \section{Elemental composition analysis:SEM-EXD}
\begin{figure}
	\centering
	\includegraphics[width=1.0\columnwidth]{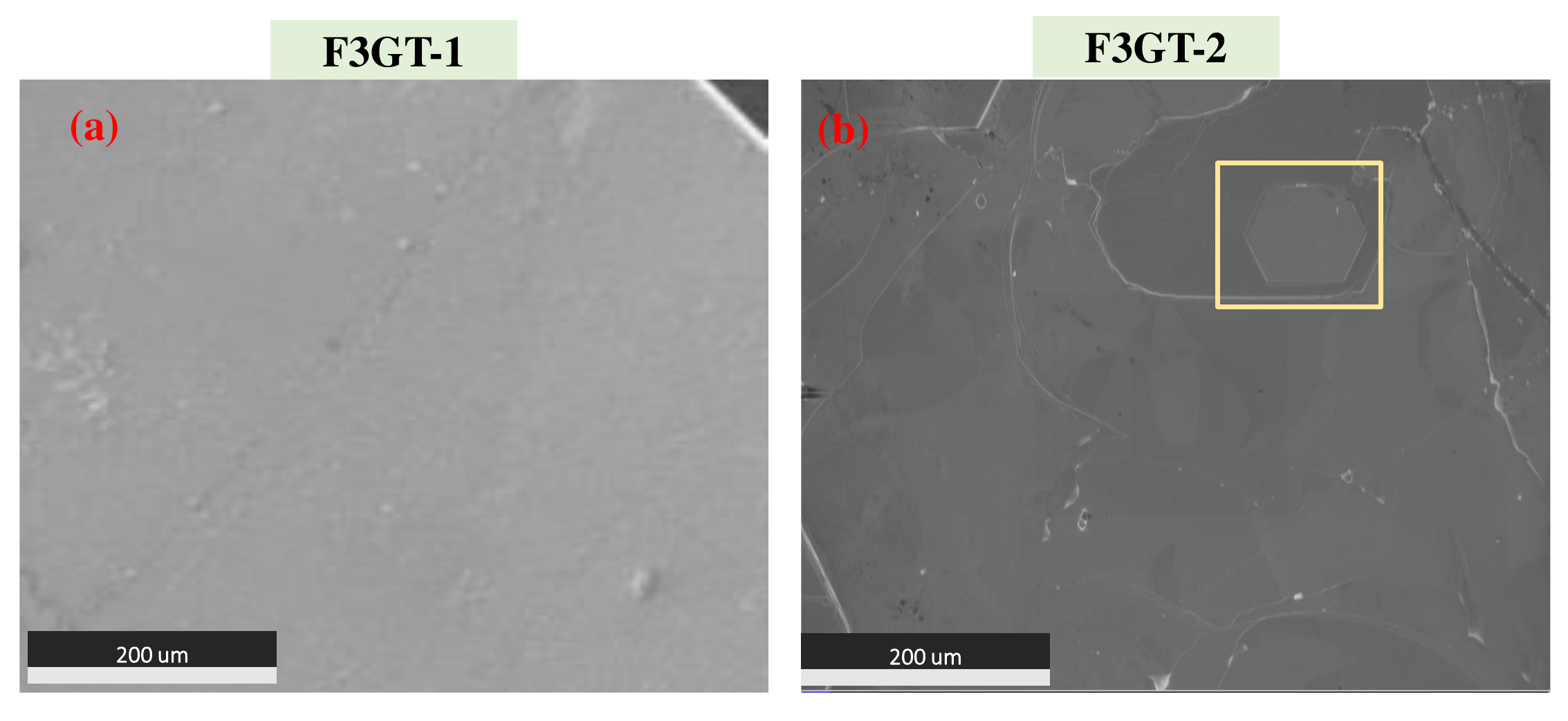}
	\caption{\textbf{Microstructural characterization.} High-resolution SEM image of (a) F3GT-1 shows a relatively smaller amount of multi-domain boundary disorder, whereas the second image (b) of F3GT-2 shows a much higher amount of multi-domain boundary  disorder.}
	\label{fig:SEM}
\end{figure}

 Figure~\textcolor{blue}{\ref{fig:SEM}(a)} describes the pure F3GT-1 phase with a relatively smaller amount of multi-domain boundary disorder, and the large area EDX (whose stoichiometry is mentioned in Section III A), was performed over the entire region enclosed by the figure whereas second figure \textcolor{blue}{\ref{fig:SEM}(b)} describes the F3GT-2 sample with a much higher amount of multi-domain boundary disorder. Here also, the large area EDX (whose stoichiometry is also mentioned in Section III A), was performed over the entire region enclosed by the
figure, whereas a separate point EDX was performed on the hexagonal-shaped flake enclosed in the yellow-colored box. It revealed a clear Ge deficiency (Ge was found to be only 2$\%$, which may have originated from the background as background has clear F3GT phase stoichiometry, i.e., molar ratio of Ge is close to 1.0). This is an indicator of FeTe phase with no Ge. Also, the flake area is very small compared to the background (which is consistent with the phase fraction $\sim$1.2$\%$ obtained from XRD analysis in section III)

\section{Resistivity}
\label{app:Resistivity}
\begin{figure}
	\centering
	\includegraphics[width=1.0\columnwidth]{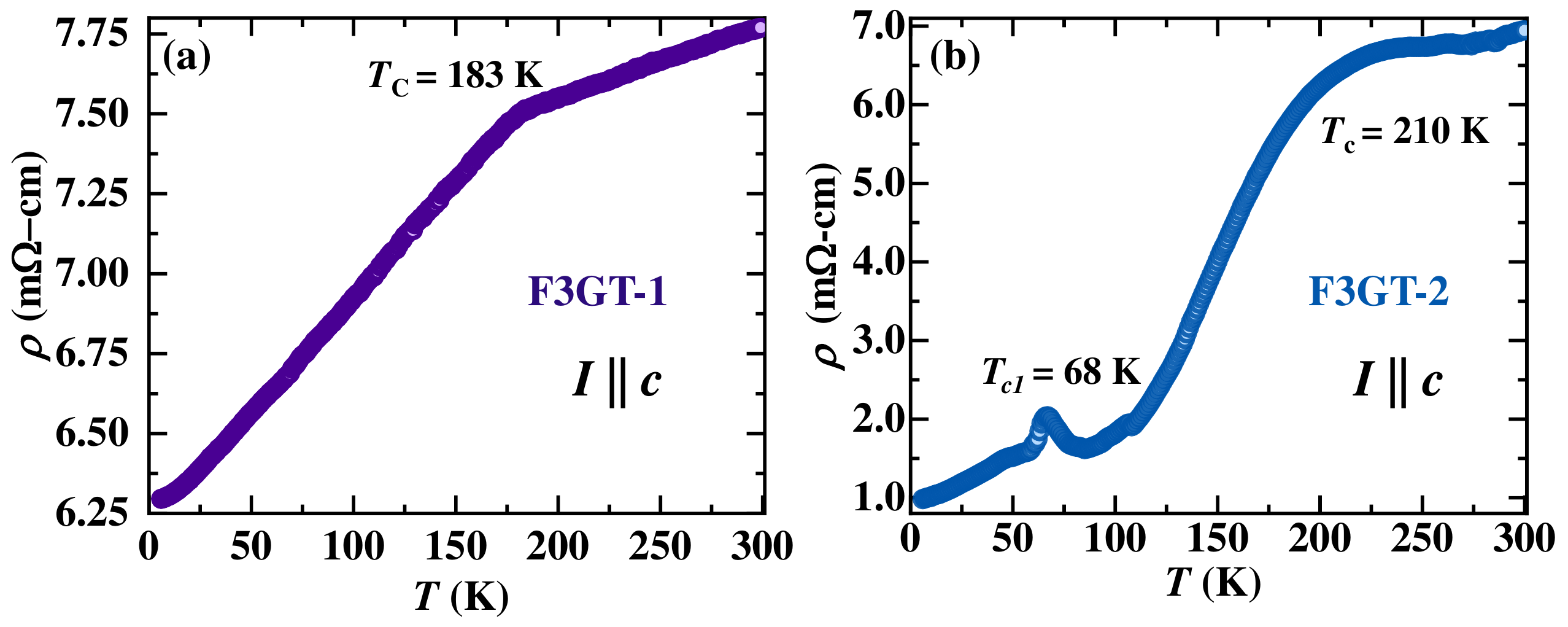}
	\caption{\textbf{Temperature-dependent electric transport measurement.} The resistivity data of both samples are parented in (a) F3GT-1 and (b) F3GT-2. The transport study shows the FM transitions of both samples at  around 200($\pm$ 10)~K consistent with earlier reports. Whereas in addition to FM transition, F3GT-2 shows the signature of AFM transition of $FeTe$ \text{at around 68 K}.  }
	\label{fig:RT}
\end{figure}

Temperature-dependent out-of-plane electrical resistivity [$\rho$(T)] of both F3GT-1 and F3GT-2 in the temperature (\textit{T}) range from 5~K-300~K are depicted in Figure~\textcolor{blue}{\ref{fig:RT}(a) and (b)} respectively. Here the observations are as follows : (i)Continuous decrease of $\rho$(T)  with decreasing \textit{T} suggests the typical metallic behavior ($d\rho/dT>$0) in F3GT-1 sample\cite{Liu2018b, Wang2017b}. A clear anomaly; a change in slope has seen observed at $T\sim$183 K, (coincide with the magnetic phase transition, described in the earlier section) due to spin disorders in the itinerant electrons\cite{Chen2013,Mao2018}. 
(ii) On the other side F3GT-2 shows similar metallic behavior. However,  F3GT-2 shows two anomalies at around 68 K and 210 K (see Figure~\textcolor{blue}{\ref{fig:RT}(b)}). The anomaly around 210 K is consistent with the magnetic data (\textit{M-T}) presented earlier. While the first anomaly (a sharp peak) at around 68~K signifies a metal-to-insulator type transition, that matches well with the previous report \cite{Mao2018,Jiang2013}.

\begin{figure}
		\includegraphics[width=1\columnwidth]{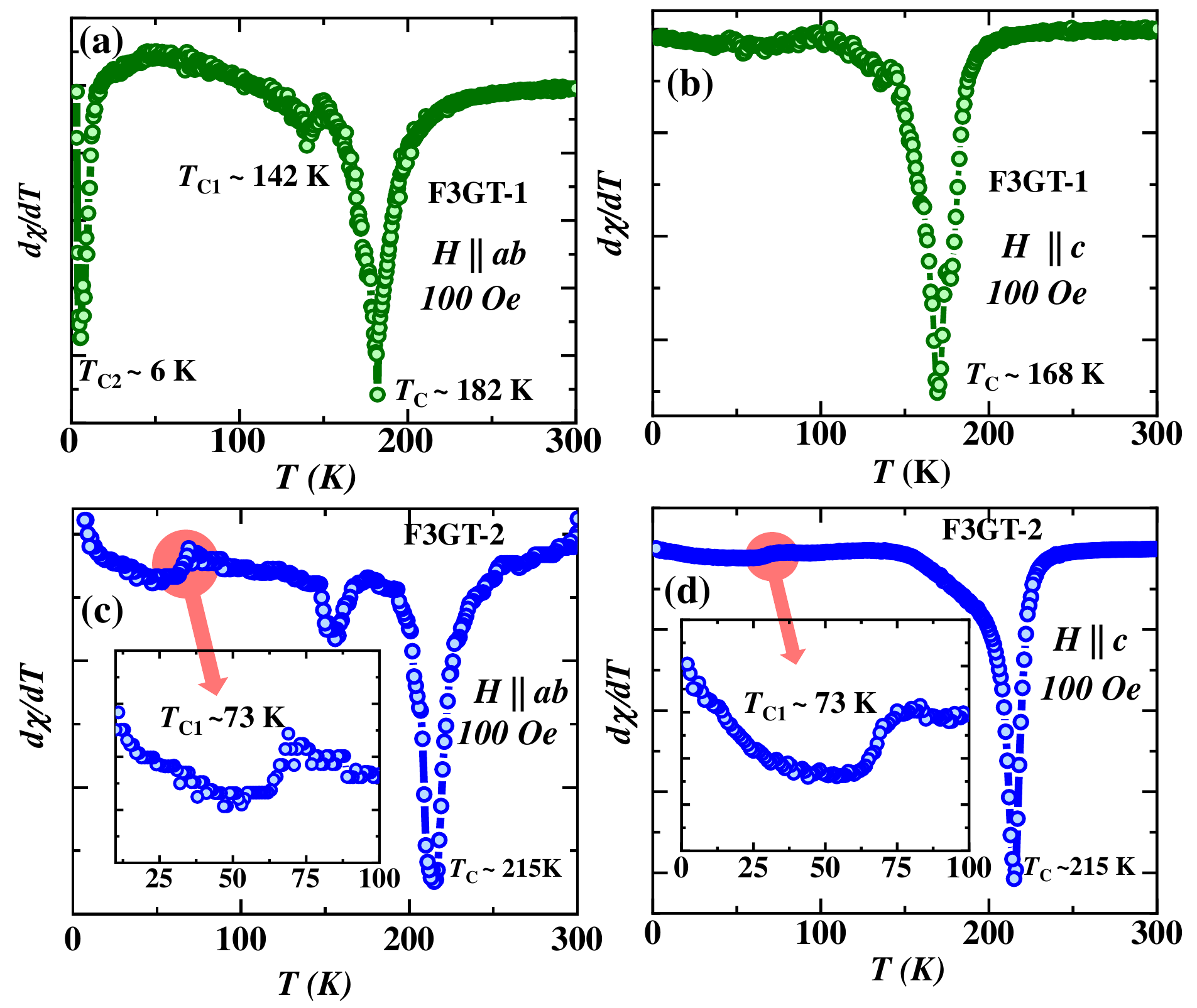}
		\caption{\textbf{The 1$^{st}$ derivative of $\chi_{dc}$(T).} The 1$^{st}$ derivative of $\chi _{dc}$(T) in both direction of F3GT-1 and F3GT-2 are represent in (a-b) \& (c-d) respectively.}
		\label{fig:dM_dT}
\end{figure}
 
\section{Magnetization of F3GT sample}
\label{app:Magnetization}	

\begin{figure}
		\includegraphics[width=0.75\columnwidth]{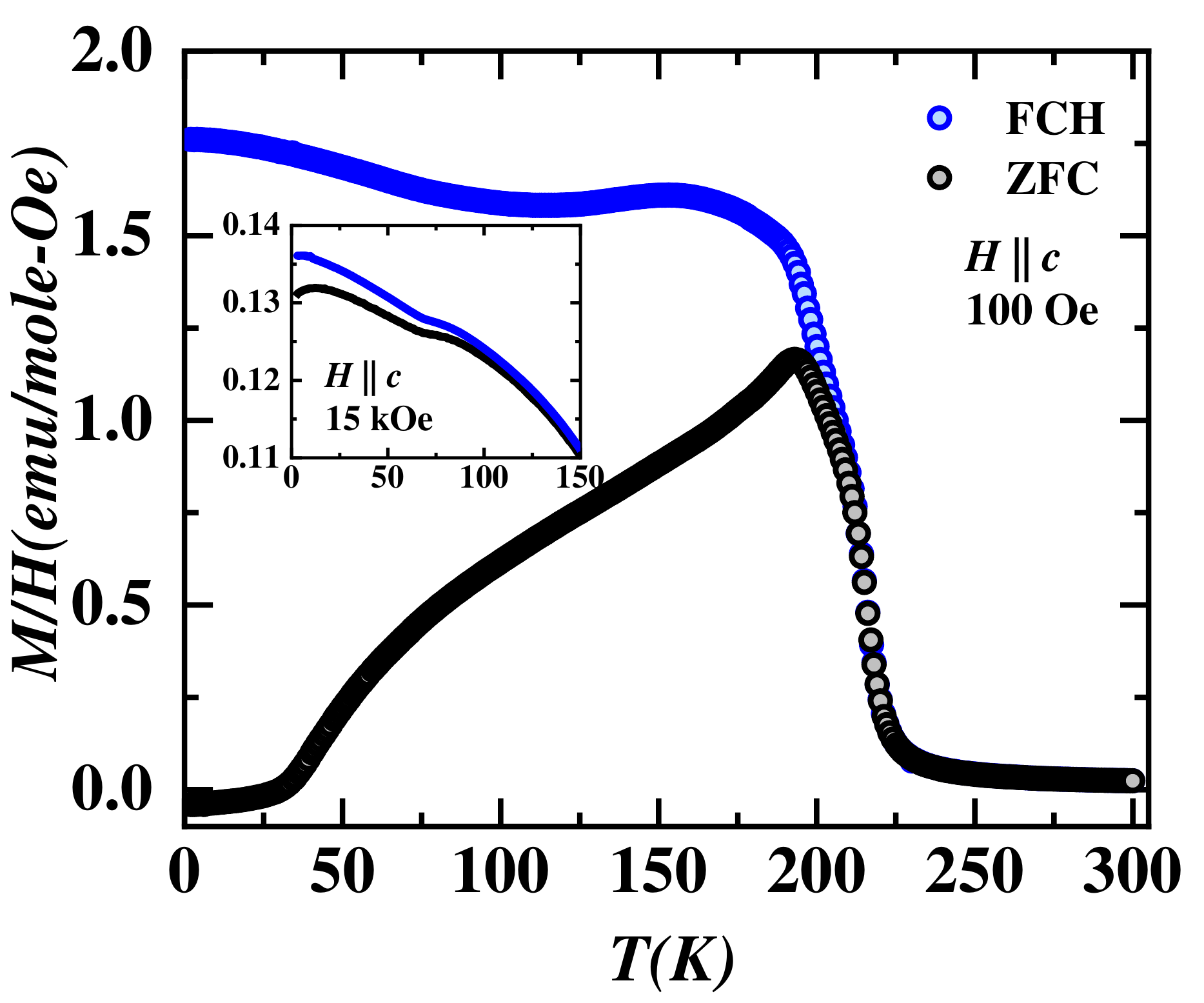}
	\caption{\textbf{Magnetization measurements.} Temperature dependence of magnetization measures with the external magnetic field \textit{H }= 0.1 kOe  and 15 kOe applied along the \textit{c} axis under zero-field-cooling (ZFC) and field-cool-heating (FCH) protocols for F3GT-2 single crystal.}
		\label{fig:MT_H}
\end{figure}

The 1$^{st}$ derivative of $\chi_{dc}$(M/H) with respect to temperature of both sample along the $ab$ plan and $c$-axis are presented in Figure~\textcolor{blue}{\ref{fig:dM_dT}}. Two prominent deep ($T_{\text{C}} \sim$ 182 K and $T_\text{C2} \sim$6 K ) and one small kink ($T_{\text{C1}}$$\sim $142 K)just below the Ferromagnetic transition are observed along $ab$ plan for F3GT-sample(see Figure~\textcolor{blue}{\ref{fig:dM_dT}(a)}). Only ferromagnetic transition at 168 K is observed along the $c$-axis of F3GT-1 sample (see Figure~\textcolor{blue}{\ref{fig:dM_dT}(b)}). On the other hand, similar kinds of features are noticed in the F3GT-2 sample (see Figure~\textcolor{blue}{\ref{fig:dM_dT}(c-d)}). At around 73 K, a small kink is observed for F3GT-2 sample, which is probably due to induces FeTe phase  in this system.
Temperature dependence of magnetization measures with the external magnetic field \textit{H }= 0.1 kOe  and 15 kOe applied along the \textit{c} axis under zero-field-cooling (ZFC) and field-cool-heating (FCH) protocols for F3GT-2 single crystal(see Figure~\textcolor{blue}{\ref{fig:MT_H}}). The bifurcation between FCH and ZFC data  increases with decreasing values of applied external magnetic field\cite{Chatterjee2020}. 

\begin{figure}
		\includegraphics[width=0.75\columnwidth]{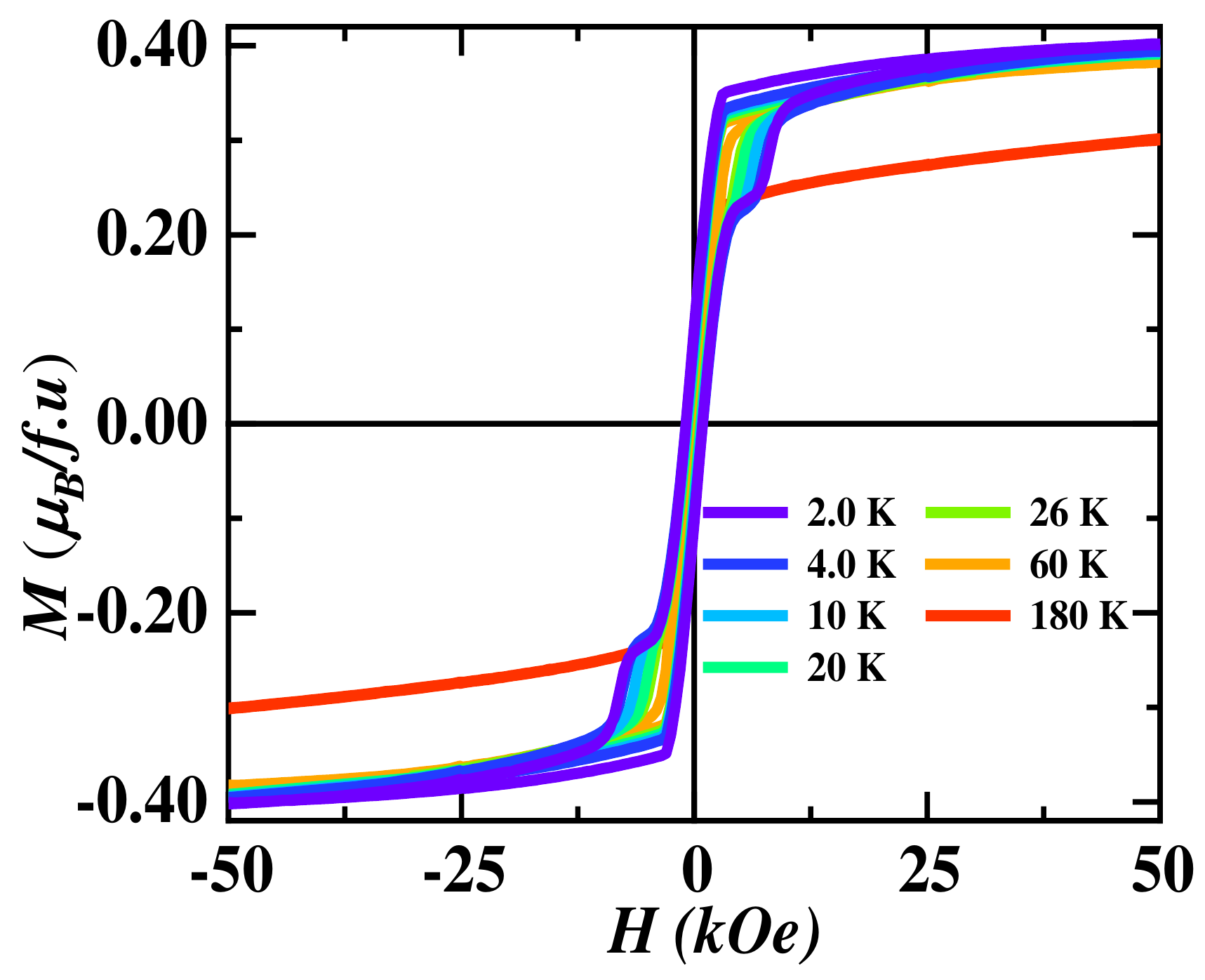}
		\caption{\textbf{Isothermal magnetization.}  Isothermal magnetization  measured  at different temperatures (\textit{T}=2, 4, 10, 20, 26, 60 and 180 K) between $\pm$ 50 kOe for F3GT-2 at various temperatures ($ H\parallel c$).}
		\label{fig:MT_H1}
\end{figure}

Figure~\textcolor{blue}{\ref{fig:MT_H1}} shows isotherm magnetization at selected temperatures (\textit{T}=2, 4, 10, 20, 26, 60 and 180 K) between $\pm$ 50 kOe for F3GT-2 at various temperatures ($ H\parallel c$). The coercive field ($H_\text{c}$) and magnetic retentivity ($M_\text{C}$) decreases with increasing temperature.

\bibliography{F3GTSubmission}

\end{document}